\documentclass[reprint,
 amsmath,amssymb,
 aps, superscriptaddress
]{revtex4-2}

\def\01{\{0,1\}}

\newcommand{\microspace}{\mspace{0.5mu}}

\def\({\left(}
\def\){\right)}

\def\<{\langle}
\def\>{\rangle}
\def \lket {\left|}
\def \rket {\right\rangle}

\newcommand{\ket}[1]{\lket\microspace #1 \microspace\rket}

\usepackage{amsmath}
\usepackage{graphicx}
\usepackage{dcolumn}
\usepackage{bm}
\usepackage{xcolor}
\usepackage{cases}
\usepackage{hyperref}

\renewcommand{\vec}{\mathbf}

\usepackage[normalem]{ulem}

\def\Rb87{^{87}\mathrm{Rb}}                             

\begin{document}
\title{Feedback cooled Bose-Einstein condensation: near and far from equilibrium}

\author{Evan P. Yamaguchi}
\affiliation{Department of Physics, University of Maryland, College Park, Maryland 20742, USA}
\author{Hilary M. Hurst}
\affiliation{Joint Quantum Institute, National Institute of Standards and Technology, and University of Maryland, Gaithersburg, Maryland, 20899, USA}
\affiliation{Department of Physics and Astronomy, San Jos\'{e} State University, San Jos\'{e}, California, 95192, USA}
\author{I.~B.~Spielman}
\affiliation{Joint Quantum Institute, National Institute of Standards and Technology, and University of Maryland, Gaithersburg, Maryland, 20899, USA}
\date{\today}

\begin{abstract}
Continuously measured interacting quantum systems almost invariably heat, causing loss of quantum coherence. Here, we study Bose-Einstein condensates (BECs) subject to repeated weak measurement of the atomic density and describe several protocols for generating a feedback signal designed to remove excitations created by measurement backaction. We use a stochastic Gross-Pitaevskii equation to model the system dynamics and find that a feedback protocol utilizing momentum dependant gain and filtering can effectively cool both 1D and 2D systems. The performance of these protocols is quantified in terms of the steady state energy, entropy, and condensed fraction. These are the first feedback cooling protocols demonstrated in 2D, and in 1D our optimal protocol reduces the equilibrium energy by more than a factor of 100 as compared with a previous cooling protocol developed using the same methodology. We also use this protocol to quench-cool 1D BECs from non-condensed highly excited states and find that they rapidly condense into a far from equilibrium state with energy orders of magnitude higher than the equilibrium ground state energy for that condensate fraction. 
We explain this in terms of the near-integrability of our 1D system, whereby efficiently cooled low momentum modes are effectively decoupled from the energetic `reservoir' of the higher momentum modes. We observe that the quench-cooled condensed states can have non-zero integer winding numbers described by quantized supercurrents.
\end{abstract}

\maketitle

\section{Introduction}

The repeated measurement of an interacting closed quantum system typically leads to runaway heating and the ultimate destruction of quantum correlations.
In some cases, the interplay of unitary evolution, measurement, and feedback can lead to physics beyond what is possible in closed quantum systems.
For example, recent developments in quantum circuits with randomized measurements combined with (sometimes random) entangling operations have shown remarkable robustness of entanglement~\cite{Li2018,Skinner2019,Cao2019,Gullans2020,Choi2020,Jian2020,Lavasani2021,Nahum2021,Alberton2021,Ippoliti2021}.
Furthermore, many quantum error correction algorithms rely on measurement of ancilla qubits followed by feedback to protect the entanglement between the constituent (logical) qubits~\cite{Shor1995,Terhal2015}. To clarify the connection between these ideas and their potential realization in steady-state many body systems, we focus on theoretically modeling Bose-Einstein condensates (BECs) in which feedback conditioned on the history of measurement outcomes provides a mechanism to retain the system in a state of low entropy, low energy, and with long-range coherence.

Many feedback schemes have been proposed to bolster quantum simulation capabilities. Ref.~\cite{Lloyd2000} showed that weak measurement combined with classical feedback can produce dynamics described by a non-linear Schrodinger equation. More recent developments have shown the utility of feedback control for creating new phase transitions, simulating spin models and preparing targeted states in optical lattices~\cite{Ivanov2020,MunozArias2020a,MunozArias2020b,Wu2021,Young2021}.
Furthermore, Refs.~\cite{hurst2019measurement, hurst2020feedback} demonstrated the utility of weak measurement and feedback for generating effective interactions in  continuum 1D BECs, with sufficient cooling to achieve a quasi-steady state. The ability to maintain the system in a quasi-steady state with long-range coherence is paramount for experimental realizations of these proposed ideas. To this end, here we describe and optimize cooling protocols for continuously monitored BECs in one and two dimensions. 

We build upon a previous framework of weak measurement and classical feedback used to control BECs developed in Refs.~\cite{hurst2019measurement, hurst2020feedback}, where the system is modeled by a stochastic Gross-Pitaevskii equation (GPE). These feedback schemes work by measuring the position resolved atomic density and applying a feedback potential that has the observed density distribution as its ground state. Because the measurement provides no information regarding the condensate phase, this procedure is imperfect. By ignoring the condensate phase, these methods neglected dynamics which are most simply described by phonons. 
The improved cooling scheme presented here (see Fig.~\ref{fig:meas-model}) uses signal filtering techniques to create feedback potentials that incorporate these effects into the feedback protocol, thereby making it more effective.

Finally, we explore the utility of weak measurement and feedback cooling in analog quantum simulation by preparing far from equilibrium Bose-condensed states.
Using optical measurements as a model for weak measurement implies that there is a resolution limit below which an observer obtains no information about the system, even in principle. We show that 1D systems with significant excitations below this scale can still be feedback-cooled to Bose condensation. In this case, the long-wavelength modes are rapidly cooled while the short wavelength modes are left relatively untouched. Because the 1D system is nearly integrable, there is negligible energy transfer between these groups of modes, yielding a low energy system weakly coupled to a highly excited reservoir. We observe that these far from equilibrium condensed states may have non-zero integer winding numbers, described here by quantized supercurrents.

This manuscript is organized as follows: In Sec.~\ref{sec:sota} we outline our simulation methods and define important metrics for exploring condensate properties at finite entropy. The rest of Sec.~\ref{sec:sota} outlines our measurement model for phase-contrast imaging, starting from the light-matter interaction Hamiltonian, and briefly reviews the concepts of signal filtering and feedback cooling essential to the protocols presented here. 

Our main results are presented in Sec.~\ref{sec:feedback-cooling}, showing feedback cooling protocols that have been significantly improved by incorporating momentum-dependent feedback. We also present the parameter space for which these protocols are optimal according to the final energy.
In Sec.~\ref{sec:far from equilibrium}, we demonstrate that in 1D our feedback cooling protocols can cool from highly-excited un-condensed states to high energy, far from equilibrium condensed states with non-zero integer winding numbers. In Sec.~\ref{sec:outlook} we conclude and present some directions for future research. Additional details regarding the measurement model are left to the appendices. 



\section{State of the art} \label{sec:sota}

\subsection{GPE with non-zero entropy} \label{sec:non-zero-S}

The usual GPE accurately describes the low temperature properties of atomic BECs, discussed in 1D and 2D in this section. Here, we use the GPE to model BECs coupled by measurement and feedback to the environment: an open quantum system.
Inspired by classical field methods~\cite{Blakie2008} to describe finite temperature systems, we begin by studying the generic equilibrium properties of a GPE when used to model a Bose system with non-zero entropy.

In the following sections, we model measurement backaction as a stochastic process introducing number fluctuations conditioned on measurement outcomes~\cite{hurst2019measurement}. This is reminiscent of stochastic noise terms used in the stochastic-projective GPE, but which are only present while the measurement process is active. Prior to developing our feedback cooling methodology, we identify the equilibrium properties of the GPE by adding a deterministic energy in the form of white noise and allowing the system to equilibrate under energy-conserving closed system evolution. We numerically model this microcanonical ensemble by simulating many trajectories with different noise realizations. Each trajectory can be characterized by its per-particle energy and momentum ground state ($k=0$) occupation probability. An ensemble of many trajectories can be assigned a condensed fraction and Von Neumann entropy.

\begin{figure}[!tbp]
	\begin{center}
    \includegraphics[width=\linewidth]{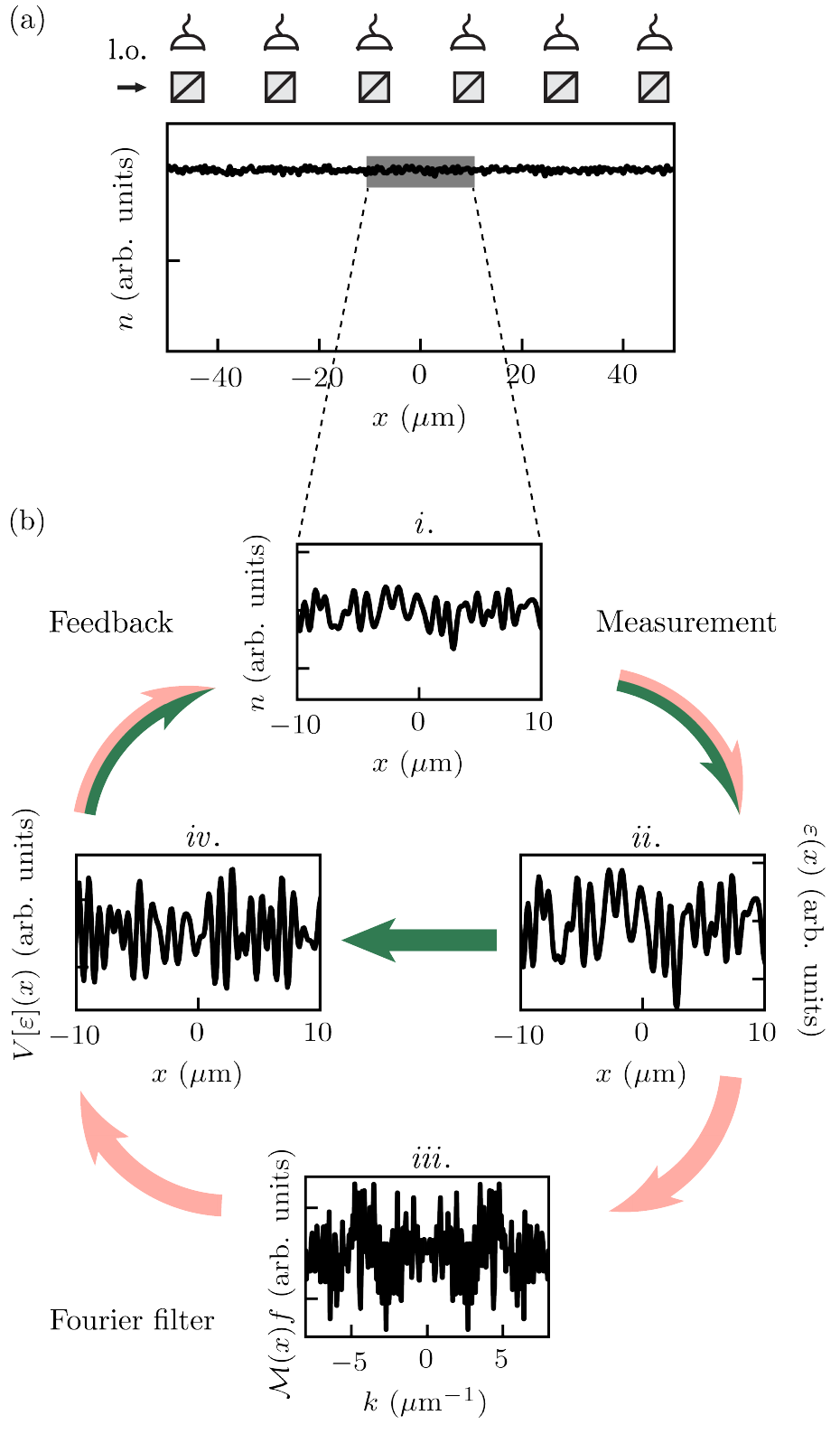}
	\end{center}
    \caption{Feedback cooling with spatiotemporal filtering.
    (a) Computed post-measurement density of a BEC weakly imaged using an off-resonant homodyne technique where l.o. indicates the local oscillator. 
    (b) Flow diagram of measurement and feedback protocols. 
    The dark green arrows trace out the basic feedback cooling protocol reviewed in Sec.~\ref{intro-feedback} and the pink arrows add the enhancements discussed in Sec. \ref{sec:feedback-cooling}. 
    (b-{\it i}) Expanded view of the post-measurement density. 
    (b-{\it ii}) The corresponding estimator $\varepsilon(x)$.
    (b-{\it iii}) Integral kernel from Eq.~\ref{kdep-estimator}.
    (b-{\it iv}) Computed potential $V[\varepsilon]$, to be fed back to the system.}
    \label{fig:meas-model}
\end{figure}

The single component GPE
\begin{align}
    i\hbar\frac{\partial}{\partial t}\Psi({\bf r},t) = \left[-\frac{\hbar^2\nabla^2}{2 m_a} + u_0|\Psi({\bf r},t)|^2 - \mu \right]\Psi({\bf r},t),
\end{align}
describes atomic BECs in the mean field limit. Here, $\Psi({\bf r},t)$ is a classical field describing $N$ particles with density $n({\bf r},t) = |\Psi({\bf r}, t)|^2$ at position ${\bf r}$ and time $t$; $\mu$ is the chemical potential; and $m_a$ is the atomic mass. Since the interaction strength $u_0$ depends on dimensionality, we will use the subscripts $u_{0, \rm 1D}$ and $u_{0, \rm 2D}$ to mark this difference where necessary.

For each trajectory, we obtain the total energy
\begin{align}
 E\left[\Psi({\bf r}, t)\right] &= \int d^D {\bf r} \left[\frac{\hbar^2}{2 m_a}|\nabla^2\Psi({\bf r},t)|^2 + \frac{u_0}{2}|\Psi({\bf r},t)|^4\right]
\end{align}
an extensive thermodynamic quantity, where the dimension $D = 1$ or $2$. An estimate of the condensed fraction can be obtained from the occupation probability of the $k=0$ mode, denoted $P(k=0)$.

We use the ensemble formed by $N_{\rm t}$ trajectories to compute additional thermodynamic quantities. The ensemble is characterized by the one body density matrix \begin{align}
\rho({\bf r}_1, {\bf r}_2) &= \frac{1}{N_{\rm t}}\sum_{j=1}^{N_{\rm t}} \Psi^*_j({\bf r}_1) \Psi_j({\bf r}_2), \label{eq:density_matrix}
\end{align}
where $j$ labels different members of the ensemble. The Von Neumann entropy $S = -\text{Tr}\left[\hat{\rho}\log\hat{\rho}\right]$ can be directly derived from $\rho({\bf r}_1, {\bf r}_2)$, giving an extensive thermodynamic quantity, where we implicitly set the Boltzmann constant $k_{\rm B} = 1$. Numerically, the entropy computed by this method is bounded from above by $S_{\rm max} = \log(N_S)$, where $N_S$ is the smaller of $N_{\rm t}$ or the number of points in our simulation grid, $N_x$. 


The Penrose-Onsager \cite{penrose1956bose} condensate number $N_{\rm c}$ is the largest eigenvalue of $\rho({\bf r}_1, {\bf r}_2)$, giving another definition of condensed fraction $f_{\rm c} = N_{\rm c} / N_{\rm t}$. A condensate is present when this eigenvalue is $\sim \mathcal{O}(N_{\rm t})$ or $f_{\rm c} \sim \mathcal{O}(1)$. This definition of condensed fraction can be directly extended to fragmented condensates where more than one eigenvalue of $\rho({\bf r}_1, {\bf r}_2)$ is $\sim \mathcal{O}(N_{\rm t})$.

{\it Simulation parameters.} We numerically simulate the the 1D and 2D GPE using split step Fourier methods~\cite{symes2016efficient}. We simulate atomic systems with density $n=10^3$ particles per micrometer, a spacing $\delta x \approx 0.1\ \mu{\rm m}$ between grid points, a time step $\delta t = 3.7\ \mu{\rm s}$, and use periodic boundary conditions.
We selected 1D and 2D interaction constants $u_{0, \rm 1D}$ $u_{0, \rm 2D}$ to give a chemical potential $\mu = h\times 165\ {\rm Hz}$, small but reasonable for $\Rb87$. The chemical potential $\mu$ determines the healing length, where $\xi = \hbar^2/2m_a\mu \approx 0.6\ \mu{\rm m}$ for the system. Energy is presented in units of $\mu$ throughout the manuscript. 


\subsection{Continuously monitored GPE}\label{subsec:formalism}

Here we briefly review our general formalism describing cold atomic gases~\cite{hurst2019measurement, hurst2020feedback, Young2021} continuously monitored with an off-resonant homodyne technique such as phase contrast imaging~\cite{andrews1997propagation}.
We show that the measurement backaction appears as a random variable in the atomic equation of motion, and we show that the dynamics of the BEC are described by stochastic wave equation.
Figure~\ref{fig:meas-model}(a) shows a schematic of the measurement apparatus for the repeated weak measurement of a $100\ \mu{\rm m}$ BEC with periodic boundary conditions in 1D.

The dispersive interaction of far detuned light and the atomic ensemble is described by the second quantized light-matter Hamiltonian
\begin{align}
\hat H_{\rm LM}&= \hbar \gamma \int \hat n({\bf r}) \otimes \hat \rho({\bf r}, z) d^{D} {\bf r}\label{eq:H3D}
\end{align}
where $\gamma$ describes the coupling strength; $D$ is the dimension; $\hat n({\bf r}) = \hat \Psi^\dagger({\bf r}) \hat \Psi({\bf r})$ measures the $D$-dimensional atomic density in terms of atomic field operators $\hat \Psi({\bf r})$; and $\hat \rho({\bf r}, z)  = \hat \phi^\dagger({\bf r}, z) \hat \phi({\bf r}, z)$ gives the optical mode occupation in terms of optical field operators $\hat \phi({\bf r}, z)$. 

We consider a pulse of light incident on our $D$-dimensional system that therefore travels in a $D+1$ dimensional space. Here, the vector $\vec{r}$ denotes the space containing the atoms and $z$ is the perpendicular dimension in which the light travels. For an incident coherent state with amplitude $\alpha$ [i.e., $\hat \phi({\bf r}, z) \ket{\alpha} = \alpha \ket{\alpha}$] and width $z = cdt$, we define a measurement strength parameter $\varphi dt^{1/2} = \gamma|\alpha| (2dt/c)^{1/2}$, a continuum analog to the expression in Ref.~\cite{hurst2019measurement}, where $c$ is the speed of light in a vacuum.

Continuous weak measurements of strength $\varphi dt^{1/2}$ output a measurement signal
\begin{align} \label{measurement-signal}
    \mathcal{M}({\bf r},t) = \<\hat{n}({\bf r},t)\> + \frac{m({\bf r}, t)}{\varphi},
\end{align}
where $\<\hat{n}({\bf r},t)\>$ is the atomic density and $m({\bf r}, t)$ is the quantum projection noise associated with the measurement. The projection noise is characterized by Fourier domain Gaussian statistics $\overline{\tilde m_{\vec k}} = 0$ and $\overline{\tilde m_{\vec k} \tilde m^*_{\vec{k}'}} = L^D\Theta(k_0-|\vec{k}|)\overline{dW_{\vec{k}} dW^*_{\vec{k}'}}/2dt^2$, $dW_{\vec{k}}$ is a Weiner increment with $\overline{dW_\vec{k}} = 0$ and  $\overline{dW_{\vec{k}} dW^*_{\vec{k}'}} = dt\delta_{\vec{k}\vec{k}'}$~\cite{anote}.

Fourier momenta greater than $k_0 = 2\pi/\lambda$ are removed from the measurement signal via the Heaviside function $\Theta$ due to the inability of the physical measurement process to resolve information on length scales smaller than $\lambda/2\pi$.

Our protocols control the condensate by applying a feedback potential $V\left[\varepsilon\right]$, generated using an estimator $\varepsilon$, defined in the following sub-section.
The measurement and quantum control protocol is described by a stochastic equation of motion for $\Psi(\vec{r}, t)$ in $D = 1$ and 2 dimensions~\cite{hurst2020feedback}:
\begin{align}
    d\Psi(\vec r) =  d\Psi(\vec r)|_\text{H} +  d\Psi(\vec r)|_\text{M} +  d\Psi(\vec r)|_\text{F},
\end{align}
where
\begin{align}
     d\Psi(\vec r)|_\text{H} &= -\frac{i}{\hbar}\left[\hat{H}_0 + u_0|\Psi(\vec r,t)|^2 - \mu\right] \Psi(\vec r) dt, \\
     d\Psi(\vec r)|_\text{M} &= \left[\varphi m(\vec r, t) - \frac{\varphi^2}{\pi}\left(\frac{k_0}{4}\right)^D~\right] \Psi(\vec r) dt, \label{update-rule} \\
     d\Psi(\vec r)|_\text{F} &= -\frac{i}{\hbar}V\left[\varepsilon\right](\vec r) \Psi(\vec r) dt, \label{feedback-term}
\end{align}
are the contributions from closed system evolution, measurement backaction, and feedback, respectively. Figure~\ref{fig:meas-model}(b-{\it i}) shows the middle $20\ \mu{\rm m}$ section of the post-measurement density $n$ of the BEC described by $\Psi(\vec{r})$ in the above equations of motion. 

\subsection{Signal Filtering} \label{intro-filtering}

Our simulations model a continuous measurement process, for which the measurement noise in Eq.~\eqref{measurement-signal} diverges as $dt\rightarrow 0$.
In reality, any physical measurement process has a non-zero detection time $\tau$; accordingly our simulations discretize time into steps of duration $\delta t \ll \tau$. 
This yields the discrete stochastic equation in Appendix~\ref{app:discrete} which quantifies the measurement strength in terms of $\kappa = \varphi (\delta t)^{1/2}$.

In general we model a physical detector's response with the convolution
\begin{align}
    \varepsilon(x, t) = \int^0_{-\infty} dt' \mathcal{M}(x, t'+t) f(t^\prime),
    \label{eqn:low-pass-integral}
\end{align}
expressed in terms of a normalized filter kernel $f(t)$.
For example, a single measurement of fixed duration $\tau$ would be described by the box filter $f_{\rm box}(t) = \Theta(t+\tau) \Theta(-t) / \tau$.

Here, we describe our measurement in terms of the low pass filter 
\begin{align}
    \tau\dot{\varepsilon}(x, t) + \varepsilon(x, t) = \mathcal{M}(x, t),
    \label{eqn:low-pass}
\end{align}
described by filter kernel $f(t) = \Theta(-t) \exp(-t / \tau) / \tau$.
We term $\varepsilon(x,t)$ as the estimator because it is a running average that estimates the true state of the system. Figure~\ref{fig:meas-model}(b-{\it ii}) displays the estimator computed following the measurement process that resulted in the post-measurement density shown in Fig.~\ref{fig:meas-model}(b-{\it i}).

This filtered measurement signal encourages us to introduce an optimal measurement strength $\kappa^*$ for which the filtered measurement result is equal to the post-measurement density $n_{|\mathcal{M}}(x)$, i.e. $\mathcal{M}(x) = n_{|{\mathcal{M}}}(x)$. 
As show in App.~\ref{app:discrete}, this leads to the optimal measurement strength
\begin{align}
\kappa^* &= \sqrt{\frac{1}{2 n_0}\frac{\delta t}{\tau}},
\end{align}
which is exact only for the box filter.
Reference~\cite{hurst2020feedback} heuristically used this expression to obtain a near-optimal measurement strength for the exponential filter, thereby coupling $\kappa$ and $\tau$.

In Sec.~\ref{sec:feedback-cooling}, our momentum dependent filtering protocols make $\tau$ a function of $k$ and thereby are not described by a single optimal measurement strength. As a result, $\kappa$ and $\tau$ are not coupled in the feedback cooling protocols presented here.

\subsection{Feedback Cooling} \label{intro-feedback}

Feedback cooling suppresses excitations, including those present prior to any measurements as well as those added by measurement backaction. Our feedback techniques apply a time-dependent feedback potential, derived from the measured density. Even in principle, this feedback cooling is limited because the measurement provides no information regarding the condensate phase, however this limitation can be mitigated with the enhanced feedback cooling techniques discussed later in Sec.~\ref{sec:feedback-cooling}.

To generate a cooling potential $V_\text{C}(x)$ given a measurement record $\mathcal{M}(x)$, we take the stationary GPE for the post-measurement wavefunction $\psi_{|\mathcal{M}}$
\begin{align}
    \mu \psi_{|\mathcal{M}} = \left[\hat{H}_0 + u_0n_{|\mathcal{M}} + V_\text{C}\right] \psi_{|\mathcal{M}} 
\end{align}
and, under the Thomas-Fermi (TF) approximation, substitute $u_0n_{|\mathcal{M}} \rightarrow g\mathcal{M}$ and compute the potential for which the $\psi_{|\mathcal{M}}$ would be the ground state of the system:
\begin{align} \label{cooling-potential}
    V_\text{C}(x) = \mu - g\mathcal{M}(x).
\end{align}
Reference~\cite{hurst2020feedback} found that $g = u_0$ is the value of gain that results in the best cooling; this was expected as a result of the Thomas-Fermi approximation.

Using an estimator of the form shown in Eq.~\ref{eqn:low-pass-integral}, the potential expression becomes
\begin{align} \label{cooling-potential}
    V_\text{C}(x) = \mu - g\varepsilon(x).
\end{align}
Figure~\ref{fig:meas-model}(b) shows this basic feedback cooling process using dark green arrows. The arrow between Figures~\ref{fig:meas-model}(b-{\it ii}) and 2(b-{\it iv}) represents the generation of a feedback potential $V[\varepsilon]$ according to Eq.~\eqref{cooling-potential}. The arrow from Figure~\ref{fig:meas-model}(b-{\it iv}) to (b-{\it i}) represents the application of $V[\varepsilon]$ to the system via the Hamiltonian term in Eq.~\eqref{feedback-term}.

In Sec.~\ref{sec:feedback-cooling}, we build on the cooling capabilities of this basic feedback protocol by implementing momentum mode-specific feedback.


\section{Enhanced Feedback Cooling} \label{sec:feedback-cooling}
The goal of feedback cooling is to depopulate high-momentum modes (i.e., phonons in our BEC), which are excited during the measurement process due to measurement backaction.
The previous cooling protocol~\cite{hurst2020feedback} outlined in Sec.~\ref{intro-feedback} is applied uniformly to all momentum modes.
In this reference, a Bogoliubov–de Gennes analysis indicated that momentum-dependent cooling protocols should give improved performance.
In this section, we introduce momentum-dependent signal filtering and gain, describe several different protocols that improve feedback cooling, and discuss the merits of each.

\subsection{Momentum-dependent Cooling \label{subsec:M-d-cooling}}

The component of the density distribution with momentum $k > k_0$ is inaccessible to the measurement process and therefore remains untouched by both measurement backaction and feedback cooling.
However, the associated modes can become populated via scattering; this contributes to heating and prevents the system from reaching a steady state~\cite{hurst2020feedback}. An optimal feedback cooling protocol should rapidly depopulate the momentum modes with $k \lesssim k_0$, so that modes with momentum $k > k_0$ cannot be populated by collisions.

Transitioning the discussion in Secs.~\ref{intro-filtering} and~\ref{intro-feedback} into Fourier space allows us to straightforwardly introduce momentum-dependent gain and filtering, e.g. $g \rightarrow g(k)$ and $\tau \rightarrow \tau(k)$.
This results in a general cooling protocol with $k$-space estimator
\begin{equation} \label{kdep-estimator}
    \varepsilon(k, t) = \int^t_\infty dt' \frac{1}{\tau(k)} \mathcal{M}(k, t')e^{-(t-t')/\tau(k)},
\end{equation}
giving the feedback potential
\begin{align} \label{kdep-gain}
    V_\text{C}(x, t) &= \mu - \mathcal{F}^{-1}\left[g(k)\varepsilon(k, t)\right](x, t),
\end{align}
where $\mathcal{F}$ and $\mathcal{F}^{-1}$ denote the Fourier transform and its inverse, and the argument of the function delineate between spatial and spectral representations.
The functions $\tau(k)$ and $g(k)$ are selected based on physical considerations discussed below. 

The time constants $\tau(k)$ in the integral Eq.~\eqref{kdep-estimator} used to compute $\varepsilon(k, t)$ allow for a momentum-dependent compromise between reduced noise (longer averaging times) and rapid response to system dynamics (shorter averaging times). Here, we used 
\begin{equation}
    \tau(k) = \tau_{\rm c} \frac{k_0}{k}
\end{equation}
where $\tau_{\rm c}$ is the filter time constant at $k_0$, the maximum accessible momentum. 
As a result, the small $k$ contributions to $\varepsilon(k, t)$ have relatively longer filter times.
We designed this function to be compatible with phonon-dynamics: a phonon of wavevector $k$ will create a density modulation whose phase changes by $1$ radian in a time $1/(c k)$.  Therefore, averaging the information over longer times would not be useful in estimating the current density of the system.

Similarly, the gain function $g(k)$ in Eq.~\eqref{kdep-gain} includes the possibility that the system's density responds to applied potentials differently on different length scales.
For example, low-$k$ phonon excitations couple to density much more weakly than high-$k$ particle-like modes; this motivated us to select a $g(k)$ to decrease with increasing $k$. This is in contrast with the optimal protocol for adiabatic feedback cooling~\cite{hurst2020feedback}, which features increasing gain at large $k$.
Our momentum-dependent gain function
\begin{equation}
    g(k) = g e^{-k^2/(2\sigma^2)}, \label{gaink}
\end{equation}
significantly attenuates the gain for momenta larger than a Gaussian width  $\sigma$.

Figure~\ref{fig:meas-model}(b) displays this enhanced feedback cooling protocol using pink arrows.
We show kernel of the estimator integral in Eq.~\eqref{kdep-estimator} in Figure~\ref{fig:meas-model}(b-{\it iii}).

\begin{figure}[!tbp]
	\begin{center}
    \includegraphics[width=\linewidth]{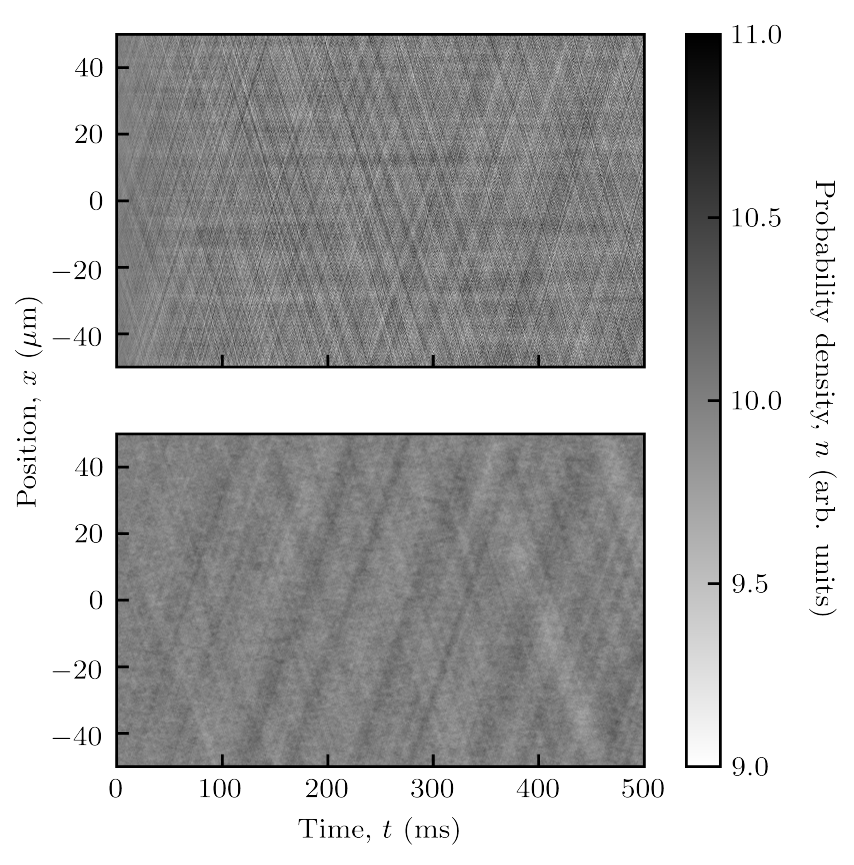}
	\end{center}
    \caption{Probability density for feedback-cooled BECs.
    The basic protocol (top) exhibits considerable structure at short length scales that is absent for the enhanced feedback cooling protocols (bottom).
    }
    \label{fig:density-comparison}
\end{figure}

\subsection{Results \label{subsec:coolingresults}}

\begin{figure*}[!tbp]
    \includegraphics[width=\linewidth]{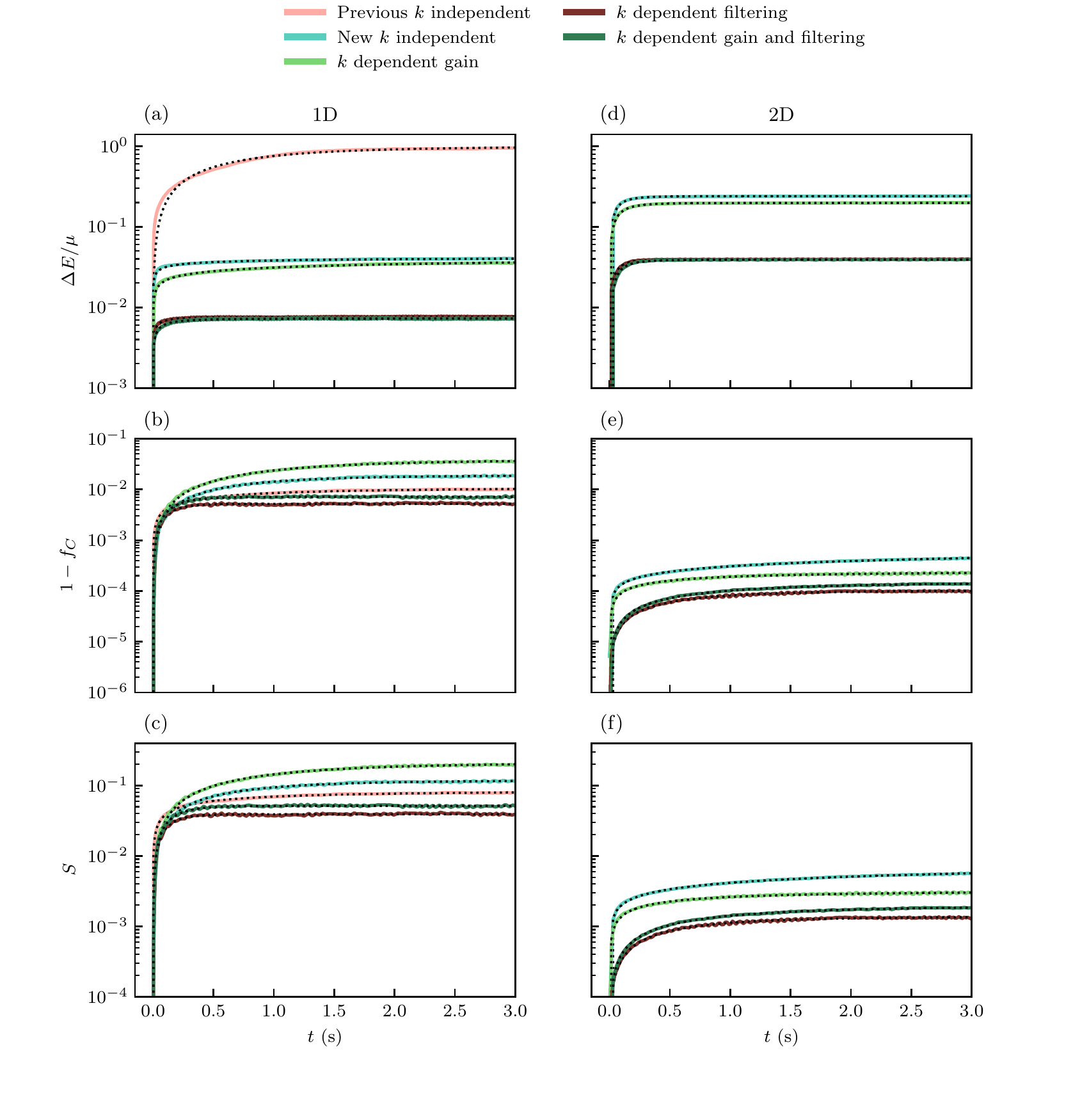}
    \caption{Comparison of feedback cooling protocols in 1D (a-c) and 2D (d-f). (a, d) Total energy difference $\Delta E$; (b, e) Un-condensed fraction $1-f_{\rm c}$, and (c) Entropy $S$.  The solid curves depict five different cooling protocols: the previous best momentum independent protocol (pink), the optimized momentum independent protocol (blue), momentum dependent gain only (light green), momentum dependent filtering only (brown), and both momentum dependent gain and filtering (dark green). The dotted black curves show fits to $\Delta E$, $1-f_{\rm c}$, and $S$; the fitting functions are discussed in Sec.~\ref{subsec:coolingresults}.
    \label{fig:protocol-comparison}}
\end{figure*}

The momentum dependent cooling protocols outlined in Sec.~\ref{subsec:M-d-cooling} cool the BEC more effectively than previous protocols according to several metrics. Figure~\ref{fig:density-comparison} displays the differences in high-frequency density noise with and without momentum dependent cooling, while Figure~\ref{fig:protocol-comparison} summarizes our results quantified in terms of total energy (top), condensed fraction (middle), and entropy (bottom) in both 1D and 2D. We also discuss the optimization of parameters $\tau_C$ and $\sigma$.

The left column of Fig.~\ref{fig:protocol-comparison} presents our 1D cooling results. In all cases the system was initialized at $t=0$ in the ground state and approached a steady state as the cooling protocols were applied. The pink curves depict the time evolution of the previous best protocol~\cite{hurst2020feedback}, which was momentum independent. We minimized the final energy by tuning the gain $g$ and filter time $\tau$ to obtain an optimal momentum independent protocol (blue), reducing the total energy by a factor of about 30. Interestingly, the momentum dependent gain protocol (light green) reduces the steady state energy, but performed worst as quantified by condensed fraction and entropy. The protocols with momentum dependent filtering reduce the total energy by an additional factor about 4 [Fig.~\ref{fig:protocol-comparison}~(a)]. The momentum dependent filter (brown) had the most significant effect on cooling across all three metrics, leading to a higher condensed fraction and lower entropy [Fig.~\ref{fig:protocol-comparison}~(b-c)]. In contrast, incorporating the momentum dependent gain parameter $g(k)$ (dark cyan) given by Eq.~\eqref{gaink} provided little advantage in cooling in terms of the total energy and lead to slightly worse outcomes in terms of the entropy and condensed fraction. 
This indicates that the physical intuition that shorter-wavelength excitations would require reduced feedback strength is either invalid, or that our Gaussian ansatz was poorly motivated.
Given the success of an overall gain derived from the Thomas-Fermi approximation, perhaps adjusting the $k$-dependent gain based upon a higher-order expansion would be better motivated.

With our previous momentum independent protocols, the system did not reach steady state---as quantified by condensed fraction, energy, and entropy---even at long times~\cite{hurst2020feedback}. By contrast, all of our metrics become time independent for the momentum dependent cooling protocols, indicating that the system is in a steady state. 

The 2D cooling results are presented in Fig.~\ref{fig:protocol-comparison}(d-f). Because there are no literature results for this type of feedback cooling in 2D, we do not plot the previous best protocol. The momentum dependent protocols reduce the total energy by a factor of slightly less than 4 compared to the 2D momentum independent protocol [Fig.~\ref{fig:protocol-comparison}~(a)], very similar to the 1D results. However, the overall energy of these protocols is about a factor of 5 higher than their 1D counterparts. By contrast, the un-condensed fraction and entropy (e,f) are about 50 times lower than those achieved by the 1D protocols. We attribute this contrast to an overall decrease in the occupation of excited modes, but with an increase in energy deriving from the increased density of states for higher momentum states. This density of states effect is even evident in individual measurements, where each measurement in 2D adds about $2.5$ times more energy than in 1D.

We quantify the overall behavior of the cooling protocols by fitting the time evolution of the energy, condensed fraction, and entropy to fitting functions of the form
\begin{equation}
E(t) = E_\infty\left[1 - \exp\left(-|t/\tau_E|^{\alpha_E}\right)\right].
\label{Eqn:efit}
\end{equation}
In this example for energy, $E_\infty$ denotes the asymptotic value of the energy, $\tau_E$ describes the nominal timescale for equilibration, and $\alpha_E$ captures the rapidity of the approach to equilibrium.

\begin{figure}[!tbp]
    \centering
    \includegraphics[width=\linewidth]{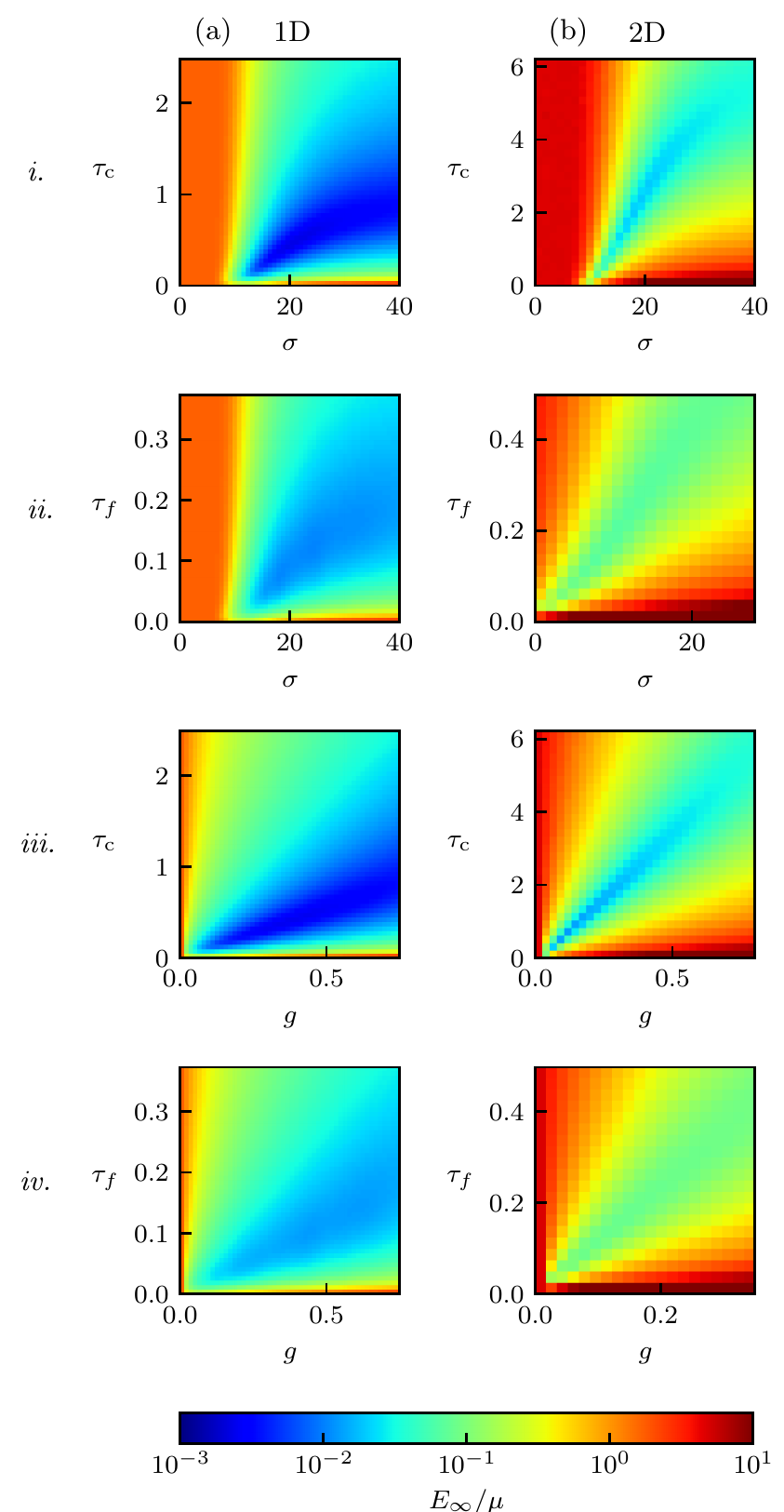}
    \caption{Final energy $E_\infty$ for two-parameter cooling protocols in both (a) 1D and (b) 2D. Numeric labels denote parameter scans for the following protocols: {\it i.} momentum dependent gain and filtering, {\it ii.} momentum dependent gain, {\it iii.} momentum dependent filtering, {\it iv.} momentum independent gain and filtering.
    }
    \label{fig:param-scan}
\end{figure}

We optimize the feedback cooling protocols by selecting the gain and filtering parameters that resulted in the lowest steady-state energy $E_\infty$. The symbols $g$ and $\tau_f$ denote the momentum independent gain and filtering protocol parameters, respectively, while $\sigma$ and $\tau_{\rm C}$ denote the momentum dependent gain and filtering protocol parameters, respectively. For each set of parameters, we simulate 16 trajectories in 1D and 5 in 2D over a period of 1 second. The results for each protocol are presented in Fig.~\ref{fig:param-scan}, where a clear local minimum emerges in the final steady-state energy. In Fig.~\ref{fig:param-scan} we imposed an upper-limit of $E_\infty/\mu \sim 10^1$; trajectories with energies beyond this limit generally failed to reach equilibrium in $1\ {\rm s}$ and are therefore not well fit by Eqn.~\eqref{Eqn:efit}. 

To reduce the parameter space for the full momentum dependent filtering and gain protocol, we selected the $k=0$ gain $g$ in Eq.~\eqref{gaink} to be to the optimal gain for momentum dependent filtering.

\section{Quench-cooling into far from equilibrium condensed states}
\label{sec:far from equilibrium}

Section~\ref{sec:feedback-cooling} presented an optimized protocol for continuously measuring a condensate while maintaining it at very low energy and entropy. In this section, we cool high energy, high entropy 1D systems with no initial condensate into final steady states with large condensed fraction and non-zero winding numbers.
These integer valued winding numbers are described by quantized supercurrents. These condensed steady states are far from equilibrium, with energy in excess of the ``critical'' energy for condensation, as discussed below. We show that this results from significant occupation of high momentum states (above the imaging cutoff $k_0$) that, because this is a nearly integrable system, are only weakly coupled to the low momentum modes accessible to feedback cooling.

\subsection{Quasi-thermal excited states}

In our simulations, we introduce ``thermal'' energy by repeatedly adding low amplitude noise over the course of $37\ {\rm ms}$ and allowing $0.5\ {\rm s}$ for the system to equilibrate.
The noise has a spectrally uniform distribution up to a cutoff $k_0 = 2\pi / \lambda$, modeling light-scattering from a near-resonant laser beam with wavelength $\lambda$.
As indicated above, we create ensembles of nominally identical trajectories (1024 trajectories in 1D and 256 in 2D), with each trajectory having an independent noise realizations. 

\begin{figure}[!tbp]
    \centering
    \includegraphics[width=\linewidth]{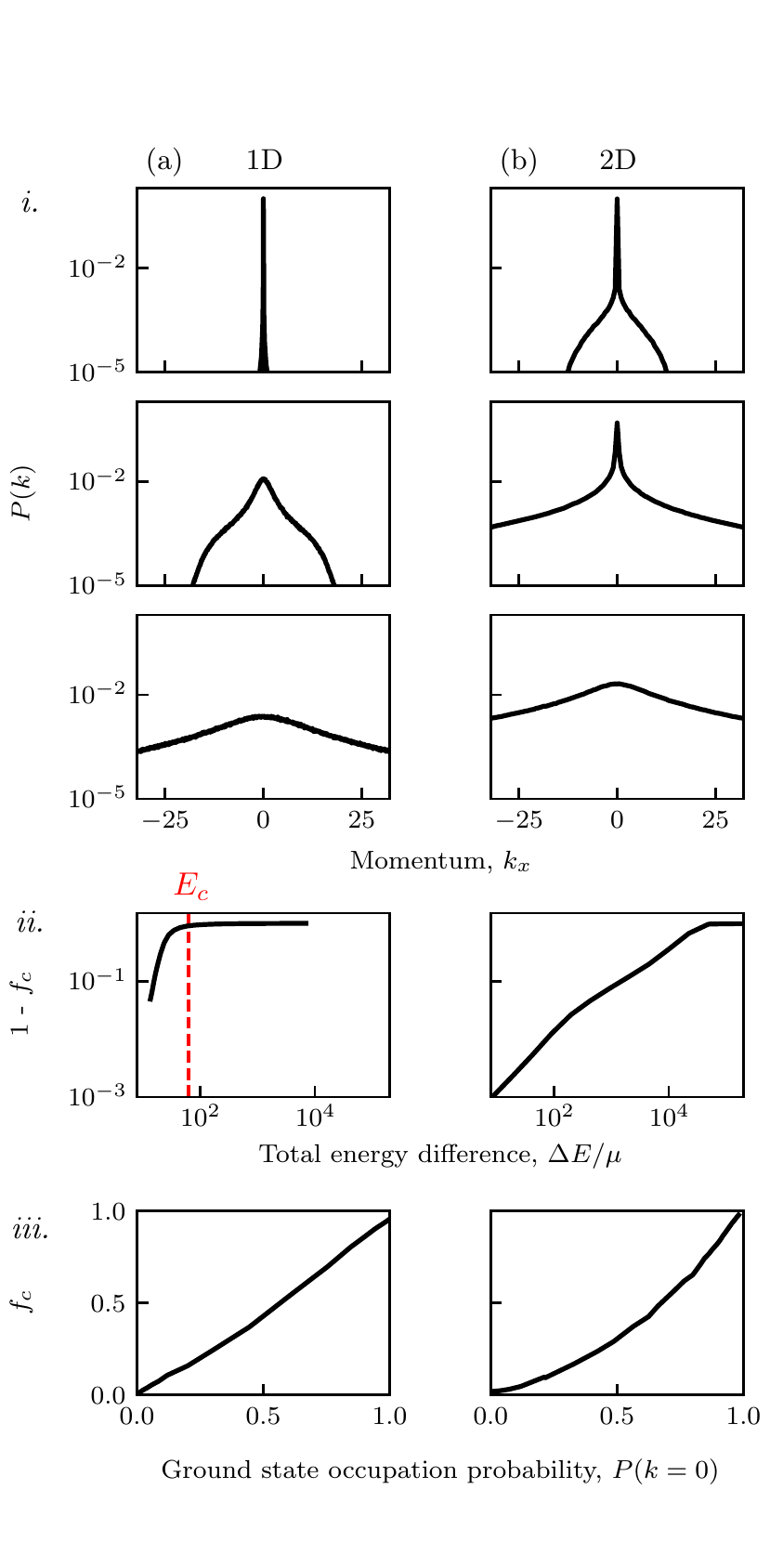}
    \caption{GPE-simulated quasi-thermal systems in (a) 1D and (b) 2D.
    ({\it i}) Momentum distributions for systems with condensed fraction $f_{\rm c} \approx 1$ (top), $0.3$ (middle), and $0$ (bottom).
    The 2D momentum distributions are projected onto the $k_x$ axis.
    ({\it ii}) Un-condensed fraction $1-f_{\rm c}$ as a function of $\Delta E$. The red dashed line denotes the transition energy $E_c$.
    ({\it iii}) Condensed fraction $f_{\rm c}$ versus the occupation probability of the $k=0$ mode $P(k=0)$.}
    \label{fig:metrics}
\end{figure}

We find that the momentum distributions in 1D did not approach a thermal equilibrium state; this is the expected behavior because the 1D GPE is an integrable model. To facilitate equilbration, we break integrability with a disorder potential. After adding the desired energy, we adiabatically ramp on this potential over $2~{\rm s}$, allow the system to equilibrate for $5~{\rm s}$ before adiabatically removing it over another $2~{\rm s}$.
This process left the total energy unchanged but increased the entropy, giving the 1D momentum distributions in Fig.~\ref{fig:metrics}(a-{\it i}), for three values of $\Delta E$ chosen to give $f_{\rm c} \approx 1$, $0.3$, and $0$. Our thermalization procedure yielded smooth distribution in the higher energy---middle and bottom---panels.
By contrast, we can see that the lowest energy case (top) has a sharp peak at the $k=0$ mode.

Figure~\ref{fig:metrics}(a-{\it ii}) shows that in 1D the un-condensed fraction $1-f_{\rm c}$ increases linearly for small $\Delta E$ and crosses over to the constant value of 1 (i.e., $f_{\rm c} = 0$) at a ``critical energy'' $E_{\rm c}$. The un-condensed fraction increases by no more than $0.1$ at energies greater than $E_{\rm c}$.
Finally, Fig.~\ref{fig:metrics}(a-{\it iii}) plots $f_{\rm c}$ as a function of the ground state occupation probability $P_{k=0}$.
We find that these quantities are nearly equal, consistent with the intuitive picture that condensate is associated with a macroscopic occupation of the ground ($k=0$) state.

In Figure~\ref{fig:metrics}(b-{\it i}), we present the 2D momentum distributions projected onto the $k_x$ axis for the same approximate $f_{\rm c}$ values as shown in 1D. As expected, we observe a sharply peaked distribution centered at $k=0$ for low energy (top), and broad smooth distributions at higher energies (middle, bottom).
No integrability-breaking disorder potential was required for thermalization in 2D.

Unlike the simple linear scaling of the un-condensed fraction with $\Delta E$ observed in 1D, Fig.~\ref{fig:metrics}(b-{\it ii}) shows that in 2D there exist regimes of different slope.
In addition, Fig.~\ref{fig:metrics}(b-{\it iii}) shows an approximately quadratic dependence of $f_{\rm c}$ on the ground state occupation $P(k=0)$. 
Although these behaviors are not a focus of our study, it is likely that as a function of increasing $\Delta E$ they result from the role of vortex anti-vortex pairs.
For example, states with a small number of vortices will have a very large $P(k=0)$, but depending on the vortex configuration can have very little wavefunction overlap leading to a small value of $f_{\rm c}$.

\subsection{Cooling from excited states}

\begin{figure}[!tbp]
    \centering
    \includegraphics[width=\linewidth]{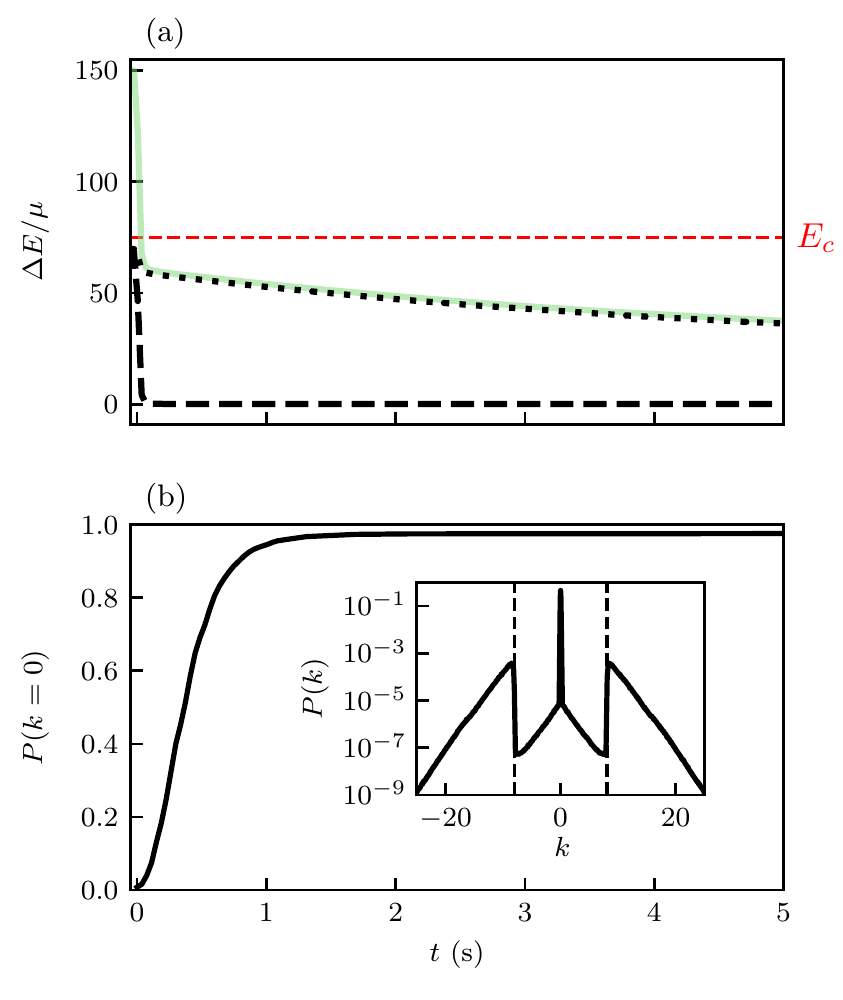}
    \caption{Time evolution following a quench from high energy.
    (a) Contributions to total energy (green), kinetic energy from momentum states above (black, dotted) and below (black, dashed) the imaging cutoff $k_0$. The red dashed line marks the transition energy $E_c$, obtained from Fig.~\ref{fig:metrics}(a-{\it ii}). 
    (b) Occupation probability of most occupied momentum mode $P (k=0)$ over the quench period.
    The inset shows the full final momentum distribution for the most energetic case in (a). The vertical dashed lines in the inset denote $|k| = k_0$.}
    \label{fig:state-characterization}
\end{figure}

Here we add energy as described in Sec.~\ref{sec:non-zero-S}, resulting in incoherent states above $E_c$, and then apply the optimized cooling protocol to generate far from equilibrium condensed states. Figure~\ref{fig:state-characterization}(a) shows $\Delta E$ as a function of time with cooling starting at $t = 0~\rm{s}$ (green) starting with an added energy $\Delta E / E_c = 2.05$.
For this amount of added thermal energy, our protocol quenches through the critical energy $E_c$ (red).

We observe that the cooling protocol is quite ineffective in energy removal for large added energy. To clarify this observation, we plot the energy divided into two parts: the kinetic energy contributions from momentum states above (dotted) and below (dashed) $k_0$ (the contribution from interaction energy is negligible).
When the cooling process begins, the kinetic energy from states with $k<k_0$ immediately plummets to a fraction of the total energy as those modes are efficiently cooled.
In stark contrast, the energy from states with $k>k_0$ slowly decreases over many seconds.
Because these modes are above $k>k_0$ they cannot be directly cooled, and instead are slowly depopulated via scattering processes.



This effect is clearly visible in the inset to Fig.~\ref{fig:state-characterization}(c) that shows the final state momentum distribution with prominent shoulders at $\pm k_0$. This distribution further displays sharp $k=0$ condensate peak surrounded by modes with occupation probabilities around $10^{-6}$. Relatively highly populated modes appear starting at $k=k_0$. We attribute the decoupling of these groups of modes and the resulting far from equilibrium state to the integrability of the 1D GPE; essentially stating that in 1D  binary collisions do not exchange momentum. 

The non-equilibrium momentum distribution in Fig.~\ref{fig:state-characterization}(c) is not observable with our weak measurements due to the finite imaging resolution set by $k_0$. However, time-of-flight imaging gives direct access to the momentum distribution, thus it could be observed experimentally. 

\begin{figure*}[!tbp]
    \centering
    \includegraphics[width=\linewidth]{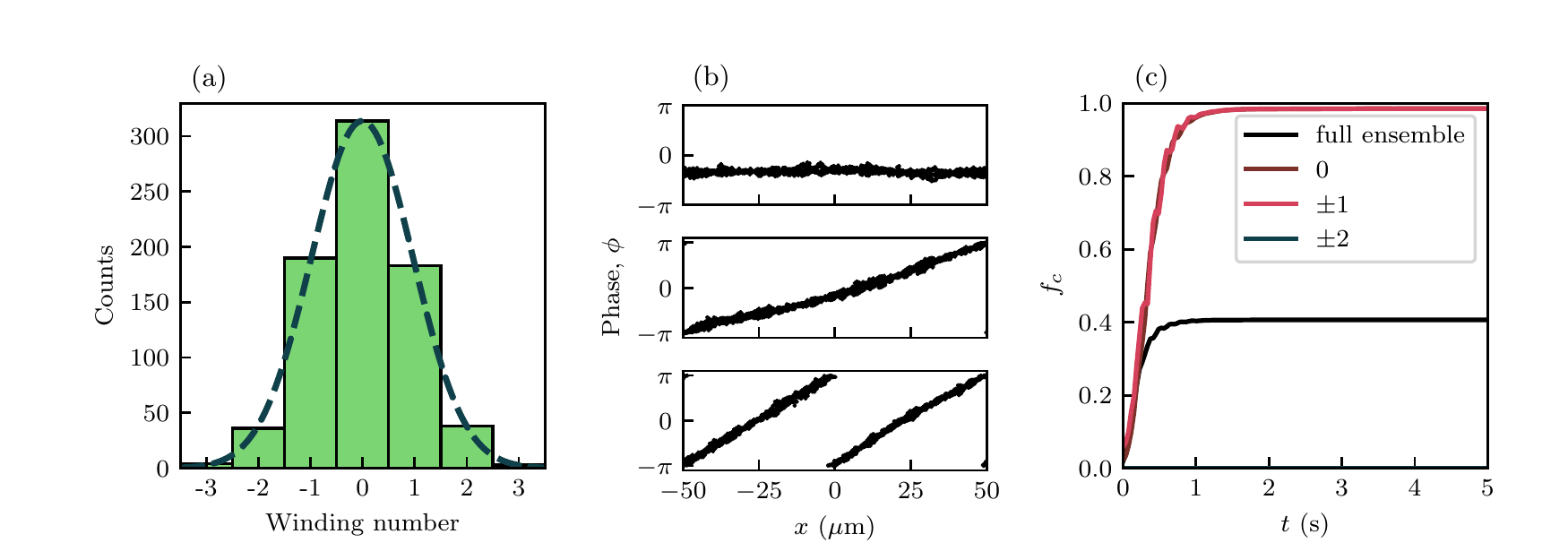}
    \caption{Winding numbers. 
    (a) Winding number distribution of a quench-cooled 768-trajectory ensemble. (b) Phase $\phi$ versus position $x$ at a single time for trajectories with winding numbers 0 (top), 1 (middle), and 2 (bottom).
    (c) condensed fraction $f_{\rm c}$ of the entire ensemble (blue) and for sub-ensembles of each winding number.
    The sub-ensembles include 36, 190, 314, 183, and 38 trajectories for winding numbers $-2$, $-1$, $0$, $1$, and $2$, respectively.}
    \label{fig:winding-numbers}
\end{figure*}

We find that our quench-cooling protocol can also result in states with integer valued winding numbers associated with quantized supercurrents, with occurrence probability plotted in Fig.~\ref{fig:winding-numbers}(a) fit to a Gaussian (with RMS width $= 0.98$).
Figure~\ref{fig:winding-numbers}(b) plots the the wavefunction phase as a function of position associated with winding numbers 0, 1 and 2. These states with non-zero winding number are still condensed, each with a single highly populated $k$-mode given by the the winding number; for example, a state with winding number 2 will be condensed into the $k=2\times 2\pi/L$ mode.

When we consider an ensemble of 768 trajectories, the Penrose-Onsager condensed fraction $f_{\rm c}$ is only about $0.3$ [Fig.~\ref{fig:winding-numbers}(c), blue], indicative of a poor condensate. However, this is an artifact of the winding number distribution in Fig.~\ref{fig:winding-numbers}(a), which leads to a formally fragmented condensate~\cite{Alon2005b}: one with many macroscopically occupied modes. We can construct sub-ensembles for each winding number, and Fig.~\ref{fig:winding-numbers}(c) shows that $f_{\rm c}$ for these modes separately have $f_{\rm c} > 0.9$.

\section{Outlook}\label{sec:outlook}

We presented a continuous weak measurement and feedback protocol that effectively removes high-energy excitations, up to the limit given by light-scattering, and preserves low-entropy BECs in both 1D and 2D. 
These results show a route toward an essential feature of any continuous feedback protocol, which is that the same quantum system can be subjected to repeated measurement without requiring an additional reservoir to remove the entropy added by the measurement process.

We furthermore demonstrated that these feedback cooling protocols can lead to new far from equilibrium condensed states that would be inaccessible in equilibrium systems. The observed separation of scales is reminiscent of condensation in pumped systems such as excition polariton condensates that can combine a non-thermal distribution of highly excited modes with a coherent condensate mode~\cite{Byrnes2014}. Interpreted in the context of traditional feedback theory, the observed momentum distributions are similar to feedback stabilized systems with Bode peaks located near the bandwidth limit.

The distribution of winding numbers in Fig.~\ref{fig:winding-numbers}(a) for these states is reminiscent of the predictions of the Kibble-Zurek mechanism~\cite{Kibble1976,Zurek1985} describing quenches through phase transitions.
In this work we did not study the width of these distributions as a function of quench speed or system size, as would be required to test if this feedback-cooling transition obeys Kibble-Zurek scaling.
If it does, a potentially interesting direction for future study would be searching for non-mean field exponents using systems with novel or non-local feedback~\cite{Young2021}.
We note that the winding number is observable, and has for example been studied in superfluid ring experiments~\cite{Eckel2014a}.

Additional work is needed to characterize analogous far from equilibrium condensed states in 2D systems.
In 2D, quantized vortices---topological defects---are analogous to the winding number in 1D, but the large parameter space given by their number and positions within the condensate make the construction of low-entropy sub-ensembles impossible and underlies the non-condensed Berezinskii-Thouless-Kosterlitz superfluid.

\begin{acknowledgments}
The authors thank S.~Guo, J.~T.~Young, and E.~Altuntas for productive discussions, and E.~Mercado and J.~T.~Young for carefully reading the manuscript.
This work was partially supported by the National Institute of Standards and Technology, and the National Science Foundation through the Quantum Leap Challenge Institute for Robust Quantum Simulation (grant OMA-2120757). HMH acknowledges the support of the NIST NRC postdoctoral program and the San Jos\'{e} State University Research, Scholarship, and Creative Activity assigned time program.
\end{acknowledgments}

\appendix

\section{Discrete and continuous measurement}\label{app:discrete}

This appendix serves to clarify the connection between the random variable $m$ defined on a continuous domain (both finite and infinite) versus a discrete domain.
Here we begin with $m({\bf x}, t)$ with unbounded ${\bf x}$, in contrast to the body of the text where we focused on a compact spatial domain of extent $L$.  

Although our mathematical description is in terms of a continuous in time random variable, any numerical simulation must be broken down into a grid with spacing $\delta x$ and extent $L$, with time steps of duration $\delta t$.
Recall that the measurement outcome at time $t$
\begin{align} 
    \mathcal{M}({\bf r},t) &= \<\hat{n}({\bf r},t)\> + \frac{m({\bf r},t)}{\varphi},
\end{align}
has a contribution from the random variable $m({\bf r},t)$ with Fourier transform $m({\bf k}, t)$.
To make the connection with finite time steps it is convenient to define $m({\bf k},t)$ in terms of its correlation functions using a temporal Dirac delta function
\begin{align*}
\overline{m({\bf k}, t) m^*({\bf k}^\prime, t^\prime)} &= \frac{1}{2}f({\bf k})\delta({\bf k}-{\bf k}^\prime)\delta(t-t^\prime)
\end{align*}
rather than the Wiener increment.
The Heaviside step function used in the manuscript is a special case of a dimensionless momentum-space apodizing function $\tilde f({\bf k})$ with $\tilde f(0)=1$ and $\tilde f({\bf k}) = \tilde f(-{\bf k})$, implying the spatial correlation function $\overline{m({\bf r}) m({\bf r}^\prime}) = f({\bf r} - {\bf r}^\prime) / 2$.

{\it Discrete temporal domain} In numerical models we consider time points spaced by $\delta t$ and generally define quantities on this grid as the average over the time interval $\delta t$, where $\delta t$ is made small enough that dynamics during this interval can be neglected.
Thus we have averages such as
\begin{align}
\mathcal{M}_t({\bf r}) &\equiv \frac{1}{\delta t} \int_{0}^{\delta t}  \mathcal{M}({\bf r}, t^\prime + t) d t^\prime.
\end{align}
By contrast, random variables such as $m_t({\bf r})$ require a slightly different average
\begin{align}
m_t({\bf r}) &\equiv \frac{1}{\delta t^{1/2}} \int_{0}^{\delta t}  m({\bf r}, t^\prime + t) dt^\prime.
\end{align}
to give the temporal correlation $\overline{m_t m_{t^\prime}} = \delta_{t,t^\prime}/2$.
This results from the usual $\sqrt{N}$ scaling of $N$ random events 
Together these lead to 
\begin{align} 
    \mathcal{M}_t({\bf r}) = \<\hat{n}_t({\bf r})\> + \frac{m_t({\bf r})}{\kappa}, \label{eq:discrete_measurement}
\end{align}
where $\kappa = \varphi\sqrt{\delta t}$.

{\it Discrete spatial domain of length $L$} For an accurate simulation, the process of discretizing the spatial domain into a grid with spacing $\delta x$ requires that $f({\bf r})$ to be nearly constant over a range of $\delta x$; in our specific case this implies that $2\pi / \delta x \gg k_0$. Applying our Fourier transform relations  (see Appendix~\ref{app:Fourier} for our conventions) again gives the discrete correlation function
\begin{align*}
\overline{m_{{\bf k}, t} m^*_{{\bf k}^\prime, t^\prime}} = \frac{L^D}{2} \tilde f ({\bf k})\delta_{{\bf k},{\bf k}^\prime}\delta_{t,t^\prime},
\end{align*}

{\it Continuous finite spatial domain of length $L$} In this case the problem is a straightforward application of the suitable inverse Fourier transform under the assumption that the extent of the domain $L$ is much larger than the characteristic width of $f({\bf r})$.
This promptly gives $\overline{m({\bf k}) m^*({\bf k}^\prime)} = L^D \tilde f({\bf k}) \delta_{{\bf k}, {\bf k}^\prime} / 2$,
the spatial part correlation function following Eq.~\eqref{measurement-signal} in the main text.

{\it Optimal measurement strength} 
Integrating the stochastic update rule over a time-interval $\delta t$ yields the updated wavefunction
\begin{align}
\Psi^\prime _j({\bf r}) &= \left[1 + \kappa m_j({\bf r}) - \frac{\kappa^2}{\pi}\left(\frac{k_0}{4}\right)^D\right] \Psi_j({\bf r}) 
\end{align}
as expected.

Following such a finite-duration measurement the condensate density is
\begin{align}
n_t^\prime({\bf r}) & \approx \<\hat{n}_t({\bf r})\> \left[ 1 + 2 \kappa m_t({\bf r}) \right]
\end{align}
at lowest order in $\kappa$.
Equating this expression with Eq.~\eqref{eq:discrete_measurement} and assuming a constant density $n_0$ allows us to define the optimal measurement strength
\begin{align}
\kappa^* &= \sqrt{\frac{1}{2 n_0}} 
\end{align}
for which the final density and measurement outcomes are equal~\cite{hurst2020feedback}.

This concept can be extended to estimators of the form
\begin{align}
    \varepsilon(x, t) = \int_0^\infty dt' \mathcal{M}(x, t+t^\prime) K(t)
    \label{eqn:low-pass-kernel}
\end{align}
where $K(t)$ is an arbitrary normalized filter kernel.
For the special case of a box kernel with width $\tau$ the estimator is
\begin{align}
\varepsilon_j(x) &= \<\hat{n}_j({\bf r})\> + \frac{m_{{\rm est},j}({\bf r})}{\kappa \sqrt{\tau}}
\end{align}
where $m_{{\rm est},j}$ is the filtered noise signal: a random variable again constructed to have variance of $1/2$.
The corresponding ideal measurement strength is
\begin{align}
\kappa^* &= \sqrt{\frac{1}{2 n_0}\frac{\delta t}{\tau}}.
\end{align}
The ideal measurement strength is reduced by a factor of $\sqrt{\delta t/\tau}$ reflecting the fact that multiple measurements of length $\delta t$ can be taken in the time $\tau$.
This yields the expected $1/\sqrt{N}$ scaling with the number of measurements taken in the time interval $\tau$.

\section{Fourier transform pairs}\label{app:Fourier}

Many expressions in this manuscript depend on the specific convention used for Fourier transform pairs. 
Our Fourier transform conventions are documented here.\\

\noindent
(CI) Continuous infinite domain.
We make the choice
\begin{align*}
f(r) &= \int \frac{dk}{2\pi} \tilde f(k) e^{i k x} &&\leftrightarrow & \tilde f(k) &= \int dx f(x) e^{-i k x}
\end{align*}
which is the defacto standard in physics, and forms the basis for the following cases.
A consequence of this choice is that a unit normalized wavefunction function $\psi(x)$ is transformed to $\tilde\psi(k)$ with a $2\pi$ norm; in other words, $dk/(2\pi)$ is the measure required to retain the correct normalization.\\

\noindent
(CF) Continuous finite domain with length $L$.
Here the pair
\begin{align*}
f(r) &= \frac{1}{L} \sum_k \tilde f(k) e^{i k x} &&\leftrightarrow & \tilde f(k) &= \int dx f(x) e^{-i k x}
\end{align*}
has a sum over $k$ that is the Riemann sum associated with the $k$ integral in the CI case, with $\delta k = 2\pi/L$.\\

\noindent
(DI) Discrete infinite domain with spacing $\delta x$.
Similar to the CF cases, the pair
\begin{align*}
f(r) &= \int \frac{dk}{2\pi} \tilde f(k) e^{i k x} &&\leftrightarrow & \tilde f(k) &= \delta x \sum_x f(x) e^{-i k x}
\end{align*}
now has a sum over $x$ associated with the $x$ integral of the CI case.\\

\noindent
(DF) Discrete finite domain consisting of $N$ sites.
In this case we approximate the differential in both sums to arrive at
\begin{align*}
f(r) &= \frac{1}{L} \sum_k \tilde f(k) e^{i k x} &&\leftrightarrow & \tilde f(k) &= \delta x \sum_x f(x) e^{-i k x}.
\end{align*}
Notice that this is not the standard textbook convention for which both expressions would have prefactors of $1/N^{1/2}$ rather than $1/L$ and $\delta x$.
Our motivation for making these choices is that the units of $f(r)$ and $\tilde f(k)$ are unchanged and the connection to continuum systems is direct and clear.

\vspace{10pt}

\section{Stochastic Wavefunction Evolution}\label{app:update}

The stochastic evolution describing the effect of measurement backaction on the BEC coherent state can be derived from the Kraus operator formalism~\cite{Kraus1974}. In this appendix, we outline the key steps of the calculation leading to Eqn.~\eqref{update-rule} in the main text. For continuous systems in space and time, the Kraus operator is
\begin{equation}
    \hat{\mathcal K}_{m} = \exp \left[-\frac{\varphi^2 dt}{2}\int d\vec{r}~\left(\delta\hat{n}(\vec{r},t) - \frac{m(\vec{r}, t)}{\varphi}\right)^2\right]
\end{equation}
where $m(\vec{r}, t)$ denotes a stochastic variable for the continuous time system, $\varphi$ is the measurement strength, and $\delta\hat{n}(\vec{r}, t) = \hat{n}(\vec{r},t) - \langle\hat{n}(\vec{r},t)\rangle$ is the density difference operator. As discussed previously, $m(\vec{r}, t)$ is defined by it's correlation function and can be expressed in terms of a Wiener process as in Sec.~\ref{subsec:formalism} or a temporal delta function as in Appendix~\ref{app:discrete}.

We begin by assuming the BEC is initially in a coherent state $|\Psi\rangle$. The action of $\delta\hat{n}(\vec{r}, t)$ on a coherent state is
\begin{equation}
    \delta\hat{n}(\vec{r}, t)|\Psi\rangle = \left[\hat{\psi}^\dagger(\vec{r}, t)\Psi(\vec{r}, t) - |\Psi(\vec{r}, t)|^2\right]|\Psi\rangle. \label{Eqn:dnmeanfield}
\end{equation}
where $\hat{\psi}(\vec{r}, t)$ indicates a bosonic field operator and the classical field $\Psi(\vec{r}, t)$ is it's mean-field counterpart. 

The updated state after measurement is formally given by $\hat{\mathcal K}_{|m}|\Psi\rangle$ which is conditioned on the measurement result via the stochastic variable $m(\vec{r}, t)$. Using a path integral formalism and starting with Eq.~\eqref{Eqn:dnmeanfield}, we define a probability distribution for the overlap of the updated state with a new bosonic coherent state $|\Psi_{|m}\rangle$, 

\begin{equation}
    \mathcal{P}\left[\Psi^*_{|m}, \Psi \right] \propto \left|\langle\Psi_{|m}|\hat{\mathcal{K}}_{|m}|\Psi\rangle\right|^2  \approx e^{-\mathcal{F}[\Psi^*_{|m}, \Psi] -\mathcal{G}[\Psi]}
    \label{eqn:probability}
\end{equation}

where $\mathcal{F}[\Psi_{|m}, \Psi]$ is a functional of the field $\Psi^*_{|m}$, and $\mathcal{G}[\Psi]$ is a functional only of $\Psi$. The state $\langle \Psi_{|m}|$ that is maximally overlapping corresponds to the saddle point, i.e. $\delta\mathcal{F}/\delta\Psi^*_{|m} = 0$. 

We consider the regime where measurement strength $\varphi$ and time increment $dt$ are both small, and expand the Kraus operator in orders of $\varphi m(\vec{r}, t)$ up to order $(\varphi m)^2$. In the following the $t$ index is suppressed for clarity. The Kraus operator in this approximation is 
\begin{widetext}
\begin{equation}
\hat{\mathcal{K}}_{m} \approx 1 + \varphi dt \int d\vec{r}~m(\vec{r})\delta\hat{n}(\vec{r}) +\frac{\varphi^2}{2}\int d\vec{r} d\vec{r}'~ \left[dt^2m(\vec{r})m(\vec{r'})-dt\delta(\vec{r-r'})\right]\delta\hat{n}(\vec{r})\delta\hat{n}(\vec{r}'). 
\end{equation}
\end{widetext}
Note that the $\delta\hat{n}(\vec{r})\delta\hat{n}(\vec{r}')$ term requires normal ordering the field operators to accurately calculate the action of $\hat{\mathcal{K}}_{|m}$ on $|\Psi\rangle$ to second order. We calculate the probability distribution using Eqns.~\eqref{Eqn:dnmeanfield}, an analogous expression for $\delta\hat{n}(\vec{r})\delta\hat{n}(\vec{r}')|\Psi\rangle$, Eqn.~\eqref{eqn:probability} and the overlap for bosonic coherent states~\cite{Altland2010} 
\begin{equation}
    |\langle\Psi_{|m}|\Psi\rangle|^2 = \exp\left[-\int d\vec{r}~\Psi^{*2}_{|m} -2\Psi^*_{|m}\Psi +\Psi^2\right].
\end{equation}

Rewriting the expression as an exponential and taking the functional derivative gives the saddle-point equation
\begin{widetext}
\begin{align}
\frac{\delta\mathcal{F}}{\delta\Psi^*_{|m}} &= 2\Psi^*_{|m}(\vec{r})\left\lbrace 1-\varphi^2\Psi(\vec{r})\int d\vec{r}'~\left[2dt^2m(\vec{r})m(\vec{r}') - dt\delta(\vec{r}-\vec{r'})\right]\Psi(\vec{r'})\right\rbrace \nonumber \\
&- 2\Psi(\vec{r})\left\lbrace 1 + \varphi dt m(\vec{r}) + \frac{\varphi^2}{2}\left[dt^2m^2(\vec{r})-dt\delta(0^+)\right] - \frac{\varphi^2}{2}\int d\vec{r}'~\left[2dt^2m(\vec{r})m(\vec{r}')-dt\delta(\vec{r}-\vec{r}')\right]|\Psi(\vec{r}')|^2 \right\rbrace
\end{align}
\end{widetext}
finally, solving the saddle point equation gives the expression for the updated coherent state
\begin{widetext}
\begin{equation}
\Psi_{|m}(\vec{r}) = \left(\frac{1 + \varphi dt m(\vec{r}) + \frac{\varphi^2}{2}\left[dt^2m^2(\vec{r})-dt\delta(0^+)\right] - \frac{\varphi^2}{2}\int d\vec{r}'~\left[2dt^2m(\vec{r})m(\vec{r}')-dt\delta(\vec{r}-\vec{r}')\right]|\Psi(\vec{r}')|^2 }{1-\varphi^2\Psi(\vec{r})\int d\vec{r}'~\left[2dt^2 m(\vec{r})m(\vec{r}') - dt\delta(\vec{r}-\vec{r'})\right]\Psi(\vec{r'})}\right)\Psi(\vec{r}).
\label{eqn:formalupdate}
\end{equation}
\end{widetext}
Where there are no complex terms, so we have written $\Psi^*_{|m} = \Psi_{|m}$. This formal expression can be further simplified on physical grounds. First, we ignore the nonolocal terms $\propto \varphi^2 \left[2 dt^2m(\vec{r})m(\vec{r}') - dt\delta(\vec{r}-\vec{r'})\right]$. These terms vanish for perfectly uncorrelated white noise (in the spatial domain) and are small as $k_0$ increases, which sets the scale of spatial noise correlations. We can also replace $m^2(\vec{r})$ by it's correlation function $\overline{m^2(\vec{r})}$; a reasonable approximation for fluctuating quantities. 

The remaining local $\varphi^2$ term can be further simplified as 
\begin{align}
     \frac{\varphi^2dt}{2} &\lim_{\vec{r}\rightarrow \vec{r}'} \left[dt\overline{m(\vec{r})m(\vec{r}')} - \delta(\vec{r}-\vec{r}') \right] \nonumber \\
     &= \frac{\varphi^2 dt}{4}\frac{1}{(2\pi)^D}\lim_{\vec{r}\rightarrow \vec{r}'}\int d\vec{k}\left[\Theta(k_0-|\vec{k}|) - 2\right] e^{i\vec{k}\cdot(\vec{r}-\vec{r}')}, \nonumber \\
     &\approx -\frac{\varphi^2}{\pi}\left(\frac{k_0}{4}\right)^D. 
\end{align}
where we use the continuum limit for the noise correlation function. In going from the second to the third line, we can make the approximation $\left[\Theta(k_0-|\vec{k}|) - 2\right] \approx -\Theta(k_0-|\vec{k}|)$, which is valid for large $k_0$. The integral can then be carried out exactly. The final result is valid for dimensions $D = 1, 2$ but could be generalized to higher dimensions. Finally, we are left with the updated state 
\begin{equation}
    \Psi_{|m}(\vec{r}) = \left[1+\varphi m(\vec{r}) -\frac{\varphi^2}{\pi}\left(\frac{k_0}{4}\right)^D\right] \Psi(\vec{r})dt.
\end{equation}

\clearpage
\bibliography{main}

\begin{thebibliography}{32}
\bibitem [{ano()}]{anote}%
  \BibitemOpen
  \href@noop {} {}\bibinfo {note} {In this work we use $\overline{~~\cdot~~}$
  to denote a statistical average and $\langle\cdot\rangle$ to denote a
  quantum-mechanical expectation value. $\tilde{m}_{\vec k}$ indicates the
  Fourier transform of $m(\vec{r})$.}\BibitemShut {Stop}%
\end{thebibliography}
\end{document}